\documentclass[aps,pra,twocolumn,superscriptaddress]{revtex4-1}

\usepackage{graphicx,enumerate,verbatim,bbold}
\usepackage{amsmath,amssymb,amsthm,float,mathrsfs}
\usepackage{dsfont}
\usepackage[colorlinks]{hyperref}
\usepackage[T1]{fontenc}
\usepackage[utf8]{inputenc}
\usepackage{lmodern}
\usepackage[caption=false]{subfig}

\begin{document}

\title{Proposal for Monitoring and Heralding Position States of Atoms\\in a One-Dimensional Waveguide}


\author{Alexandre Roulet}
\affiliation{Centre for Quantum Technologies, National University of Singapore, 3 Science Drive 2, Singapore 117543, Singapore}

\author{Colin Teo}
\affiliation{Centre for Quantum Technologies, National University of Singapore, 3 Science Drive 2, Singapore 117543, Singapore}

\author{Huy Nguyen Le}
\affiliation{Centre for Quantum Technologies, National University of Singapore, 3 Science Drive 2, Singapore 117543, Singapore}

\author{Valerio Scarani}
\affiliation{Centre for Quantum Technologies, National University of Singapore, 3 Science Drive 2, Singapore 117543, Singapore}
\affiliation{Department of Physics, National University of Singapore, 2 Science Drive 3, Singapore 117542, Singapore}

\date{\today}

\begin{abstract}
We study single-photon transport in a single-mode waveguide, in which two-level atoms in spatial superpositions are embedded. We find that the transmission of the photon can be used as a non-destructive probe for the coherence of the superposition. Under certain trapping configurations, the protocol proposed is shown to be independent of which wells the atoms are initially trapped in. Furthermore, we show that by looking at the transmission in a suitable regime, a maximally entangled state can be post-selected from an unknown superposition.
\end{abstract}


\maketitle

\section{Introduction}

The coherent transport of a single photon in a one-dimensional waveguide with an embedded two-level atom is characterized by interferences between the wave function of the spontaneously emitted photon and the incident wave. This conceptually simple set-up, studied almost a decade ago \cite{Shen2005}, leads to full reflection of the photon at resonance. Recently, quantum scattering with two or more photons and two- or multi-level atoms have been investigated. A number of phenomena have been predicted, such as transmission peaks near resonance \cite{Tsoi2008}, bound states of photons mediated by the scatterer \cite{Shen2007,Witthaut2010,Zheng2012,Martens2013} resulting in a partial Hong-Ou-Mandel effect \cite{Longo2012} or polarization control of a single photon \cite{Tsoi2009}.
In most of these studies, however, the main goal is to describe how a well-defined atomic configuration modifies the coherent transport of light in the waveguide.

Here, we reverse the paradigm and focus on how to \textit{create quantum spatial configurations of the atoms and assess them}, based on the modified transmission of a single photon. We note that this approach has already been adopted recently in \cite{Chang2013}, where self-organization of the atoms is probed through the transmission spectra, in the absence of externally imposed trapping potentials. Here, we propose to use trapping potentials in order to create atomic spatial superpositions. These superpositions have been studied as an analogue of two-level \cite{Mompart2003,*Eckert2005} and three-level \cite{Eckert2004,*Eckert2006} optics, as well as in quantum dot systems \cite{Greentree2004}, and observed in spin-dependent optical lattice potentials \cite{Mandel2003PRL,*Mandel2003Nature}.

\begin{figure}
\subfloat{
\includegraphics[width=0.23\textwidth]{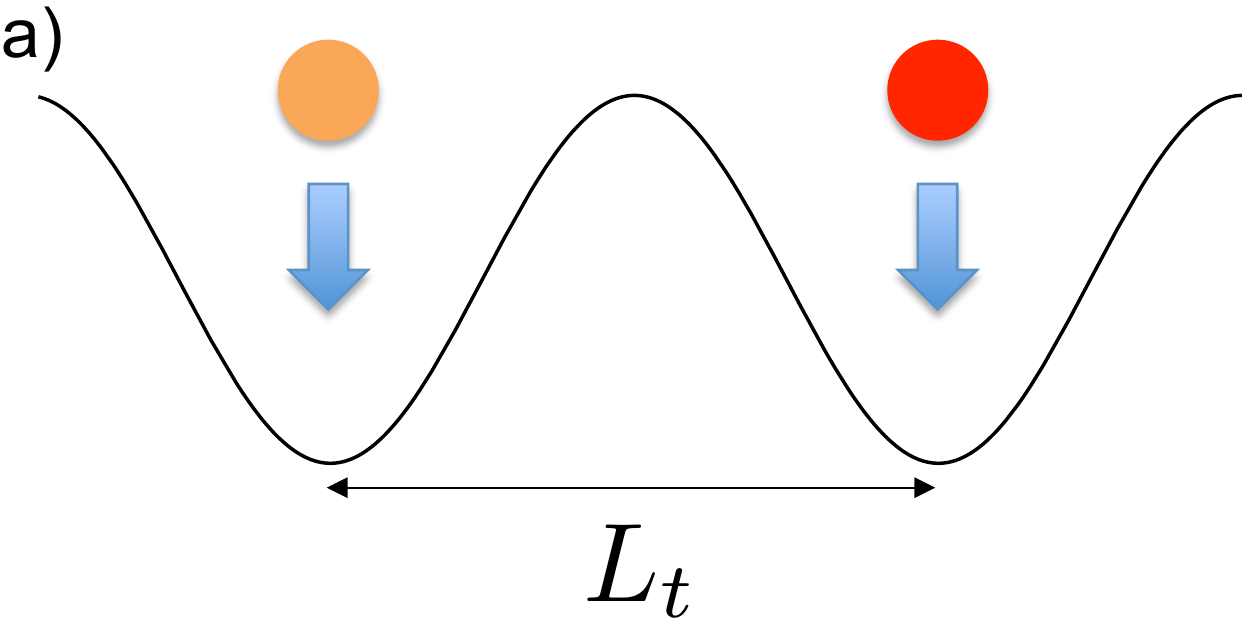}\label{fig:potentialIni}
}
\subfloat{
\includegraphics[width=0.23\textwidth]{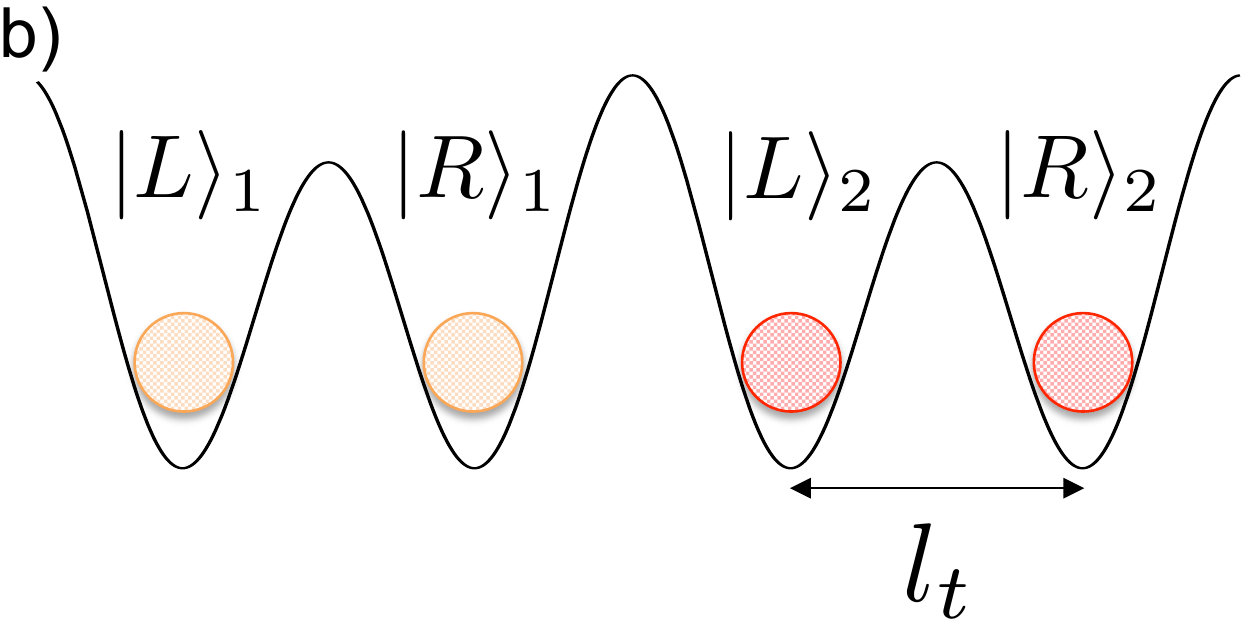}\label{fig:potentialFin}
}\\
\subfloat{
\includegraphics[width=0.47\textwidth]{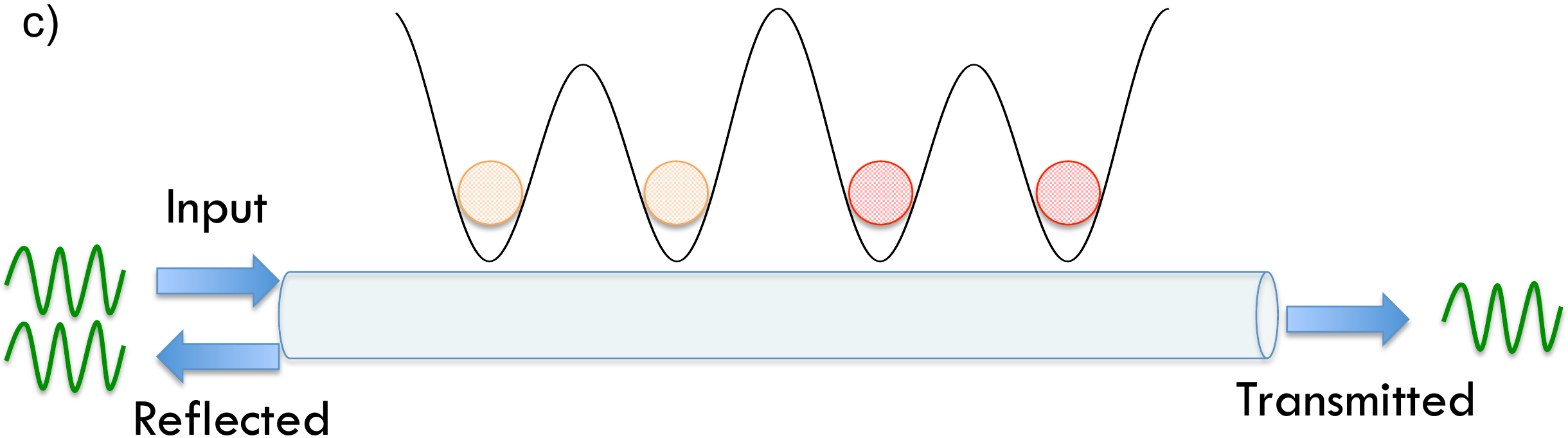}\label{fig:probingProcess}
}
\caption{\label{fig:1}(color online).\quad Trapping and probing schemes in the case of two atoms in adjacent wells. (a) Simple-well configuration ($A=0$) with localized atoms (dark colors). (b) Double-well configuration ($A\ne 0$) with atoms in a spatial superposition (light colors). (c) Probing of the atomic spatial state through the photon transmission and reflection spectra.}
\end{figure}

There are several ways of trapping atoms in a waveguide set-up \cite{Hung2013}, for instance either within hollow-core optical fibers \cite{Christensen2008,Bajcsy2011} or in the evanescent field of fiber-taper waveguides \cite{Vetsch2010,Goban2012}. Based upon the different trapping schemes designed and demonstrated, we assume in this theoretical study a periodic trapping potential of the form
\begin{equation}\label{eq:trappingPot}
	V(x)=\sin^2(k_t x)+A^2 \cos^2(2k_t x),
\end{equation}
where $k_t$ is the wave vector of the first trapping laser and $A$ is the tunable amplitude of the second trapping laser (as illustrated in Fig.\,\ref{fig:1}). However, the results do not depend critically on this expression, unless specifically mentioned. In order to create a coherent spatial superposition, we propose to start from a simple-well configuration, where the atoms are in a definite well separated by $L_t=n\pi / k_t$, with $n\in\mathbb{N}^*$ (see Fig.\,\ref{fig:potentialIni}). By adiabatically increasing $A$, a double-well potential is reached, separated by $l_t$, such that the tunnelling probability \cite{Eckert2004} becomes negligible and $L_t\approx 2nl_t$ (as shown in Fig.\,\ref{fig:potentialFin}). This amounts, for any atom $j$ within a double-well, to creating the coherent atomic spatial superposition $|\phi\rangle_j = c_L |L\rangle_j + c_R |R\rangle_j$, where $|L\rangle$ ($|R\rangle$) corresponds to the left (right) side of the double-well and $c_L,c_R\in\mathbb{C}$ are \textit{a priori} unknown. In the rest of our study, we consider the tractable case of two atoms. Specifically, we use the modified transmission of the photon to certify that each atom is in the desired coherent superposition (this requires at least two atoms, since the probability of transmission is position-independent in the case of a single atom). Further, we show how to prepare heralded entanglement of the two atoms by choosing the suitable frequency of the probe light.

\section{Model}
Consider then $N=2$ identical two-level atoms with resonance frequency $\omega_A$, separated by distance $d\equiv |x_2-x_1|$, embedded in an infinitely long waveguide with negligible lateral loss. $|g\rangle_j$ and $|e\rangle_j$ are, respectively, the ground and excited states of atom $j$ and we assume that $\omega_A$ is much larger than the cutoff frequency of the waveguide, such that the longitudinal wave number $k$ of the photon modes obeys the linearized dispersion relation $\omega=c|k|$ \cite{WaveguideBook}. The dipole Hamiltonian describing the interaction between the atoms and propagating photons, under rotating wave approximation, is given by \cite{Domokos2002}
\begin{eqnarray}\label{eq:Hamiltonian}
	&&\hat{H}= \sum_{j=1}^2 \hbar\omega_A |e\rangle_j\langle e|+\int_{0}^\infty\!\mathrm{d} \omega\, \hbar\omega \left(\hat{a}_\omega^\dagger \hat{a}_\omega+\hat{b}_\omega^\dagger \hat{b}_\omega\right)\\
	& &-i\hbar\sum_{j=1}^2\int_{0}^\infty\!\mathrm{d} \omega\, g_\omega \left[\hat{\sigma}^j_+ \left(\hat{a}_\omega e^{i\omega x_j/c}+\hat{b}_\omega e^{-i\omega x_j/c} \right)-\mathrm{H.c.} \right], \nonumber
\end{eqnarray}
where $g_\omega$ is the coupling constant, the atomic raising ladder operator is defined as $\hat{\sigma}^j_+\equiv |e\rangle_j\langle g|$, and $\hat{a}_\omega$ ($\hat{b}_\omega$) is the annihilation operator of the forward- (backward-) propagating photon mode.

We consider a single-photon pulse centered around the frequency $\omega_0$ incident from the left (see Fig.\,\ref{fig:probingProcess}), while the atoms are initially in the ground state. Assuming that the atoms only interact with the modes of the waveguide, we can then obtain the reflection and transmission amplitudes of the photon from the average number of reflected photons $N_{\text{ref}}(t)=\int_0^\infty\!\mathrm{d} \omega\, \langle \psi (t)| \hat{b}_\omega^\dagger \hat{b}_\omega |\psi (t)\rangle$, where $|\psi (t)\rangle$ is the state of the system at time $t$ (see Appendix \ref{sec:appA}). In the limit where the input-pulse bandwidth $\Omega$ is much smaller than the interaction strength $\gamma\equiv 2\pi g_{\omega_A}^2$, our results coincide with Shen and Fan's calculation \cite{Shen2005} which was derived for a monochromatic input pulse. For simplicity, we will focus on this regime $\Omega\ll\gamma$ for the rest of the paper (the case of arbitrary $\Omega/\gamma$ is presented in Appendix \ref{sec:appA}). Then, the transmission coefficient $|t_d|^2$ (and reflection coefficient $|r_d|^2=1-|t_d|^2$) reads \cite{Tsoi2008}
\begin{equation}\label{eq:transCoeffSF}
	|t_d|^2=\frac{(\Delta /\gamma)^4}{\left( (\Delta /\gamma)^2-1+\cos(2\theta_d) \right)^2+\left( 2(\Delta /\gamma)+\sin(2\theta_d) \right)^2},
\end{equation}
where $\Delta\equiv\omega_0-\omega_A$ is the detuning and $\theta_d\equiv \omega_0 d/c$ is introduced for further simplifications.

\begin{figure}
\includegraphics[width=0.46\textwidth]{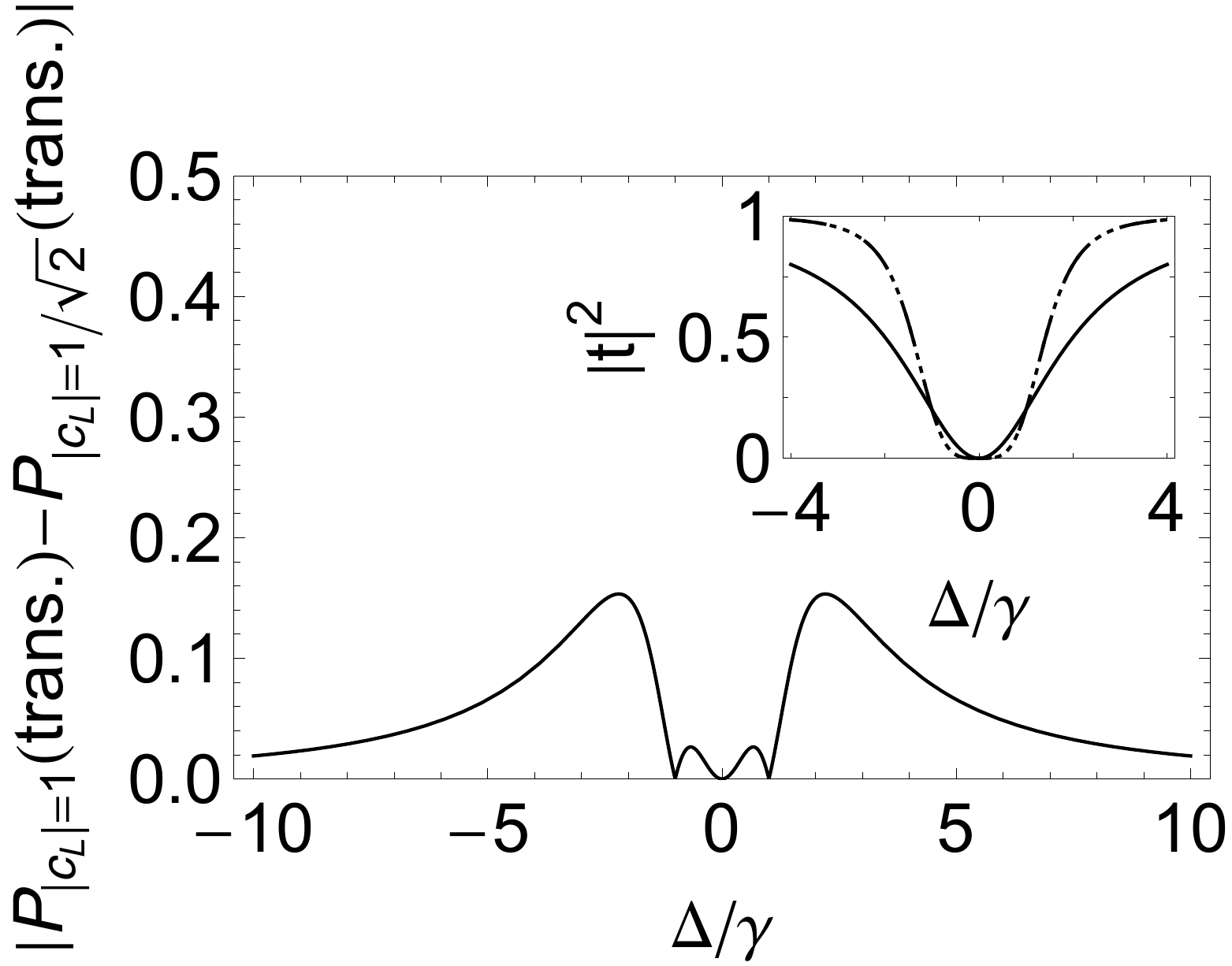}
\caption{\label{fig:partialTomo}Absolute difference between the two extreme probability distributions that can be obtained for $kl_t=\pi/2$. Note that it is upper bounded by 1/2. Inset: transmission coefficient for the different configurations. The solid, dashed, and dotted line represent respectively the $d^0$, $d^+$, and $d^-$ configurations. Note that the last two coincide.}
\end{figure}

\section{Monitoring the position state}
In the case of two atoms, the atomic spatial-superposition state
\begin{eqnarray}\label{eq:atomicSupState}
	|\phi\rangle&=&(c_L |L\rangle_1 + c_R |R\rangle_1)\otimes(c_L |L\rangle_2 + c_R |R\rangle_2)\nonumber\\
	&=&c_L^2|LL\rangle+c_R^2|RR\rangle+c_Lc_R(|LR\rangle+|RL\rangle)
\end{eqnarray}
corresponds to four different atomic configurations in the interaction basis. More precisely, the first two terms $|LL\rangle$ and $|RR\rangle$ correspond to an interatomic distance $d^0\equiv L_t$ while $|LR\rangle$ and $|RL\rangle$ correspond, respectively, to $d^+\equiv L_t+l_t$ and $d^-\equiv L_t-l_t$. From the reduced density matrix of the photon, together with the normalization condition $|c_L|^2+|c_R|^2=1$, we obtain the probability of transmission as follows
\begin{eqnarray}\label{eq:probTrans}
	P_{|\phi\rangle}(\text{trans.})&=&|c_Lc_R|^2\left[|t_{d^+}|^2+|t_{d^-}|^2-2|t_{d^0}|^2\right]+|t_{d^0}|^2.\nonumber\\
\end{eqnarray}
One can easily check that $|c_L|=0$ and $|c_L|=1$ lead to the same probability $|t_{d^0}|^2$. Therefore the states $|LL\rangle$ and $|RR\rangle$ are indistinguishable when probing only the photon transmission. Indeed, the only difference between these two configurations is a phase shift for the reflected photon. This implies that the probability of transmission (\ref{eq:probTrans}) is not invertible as a function of $|c_L|$, which precludes the possibility of performing a complete tomography. Note, however, that this is already transparent at the level of the atomic state (\ref{eq:atomicSupState}), which is symmetric under the interchange $(c_L,c_R) \rightarrow (c_R,c_L)$ for any measurement incapable of distinguishing $|LL\rangle$ from $|RR\rangle$. Hence, as long as $|t_{d^+}|^2+|t_{d^-}|^2-2|t_{d^0}|^2\ne 0$, the modulus $|c_L|$ can readily be obtained, up to this symmetry, from the transmission spectra (\ref{eq:probTrans}) and using the identity $|c_L|^2=\frac{1\pm \sqrt{1-4|c_Lc_R|^2}}{2}$.

Thus, the modulus of the spatial-superposition weights has been obtained through the transmission spectra. However, this is not sufficient in order to certify coherence. Indeed, such a transmission spectra could very well be explained by a statistical mixture corresponding to the same weights, for instance 
\begin{eqnarray}
	\hat{\rho}&=&|c_L|^4|LL\rangle\langle LL|+|c_R|^4|RR\rangle\langle RR|\nonumber\\&&+|c_Lc_R|^2(|LR\rangle\langle LR|+|RL\rangle\langle RL|).
\end{eqnarray}
To get further information on the coherence of the superposition, we need access to the relative phase between the coefficients $c_L$ and $c_R$. This is allowed by applying a Hadamard gate on both atoms prior to sending the probe photon. Specifically, the Hadamard gate performs the transformations $|L\rangle \rightarrow (|L\rangle+|R\rangle)/\sqrt{2}$ and $|R\rangle \rightarrow (|L\rangle-|R\rangle)/\sqrt{2}$. While this procedure is implemented with the help of $\pi /2$ Rabi pulses in the context of cavity QED \cite{Hagley1997}, we consider here the adiabatic lowering of the tunable amplitude $A$ such that each atom can coherently tunnel between the left ($|L\rangle$) and right ($|R\rangle$) wells of the double-well configuration. This scheme is formally identical to that proposed in \cite{Mompart2003,*Eckert2005}, up to mapping the variation in distance of the wells to a variation in barrier height in the present case. We emphasize that applying the gate is a controlled coherent process. By never suppressing the barrier completely, we ensure that the harmonic approximation remains valid, so that the evolution is governed by Rabi-type oscillations between the states $|L\rangle$ and $|R\rangle$, in contrast to the preparation procedure where the barrier is initially absent. Following these principles, we can then extract $|\cos(\varphi)|$ from the transmission spectra, where $\varphi$ is the relative phase between $c_L$ and $c_R$, and the discussed symmetry now translates into a sign incertitude. Hence we find up to four different relative phases, namely $\pm\varphi \pmod{\pi}$.

The procedure described above is based upon full knowledge of the system prior to creation of the spatial superposition. More specifically, it assumes that we know $n$, \textit{i.e} in which traps the atoms were initially loaded. However, this knowledge is not required if one adjusts the trap such that $kl_t=\pi/2 \pmod{\pi}$. Indeed, in this case one gets $\theta_{d^0}=0 \pmod{\pi}$ and $\theta_{d^\pm}=\pi/2 \pmod{\pi}$, such that the corresponding transmission spectra are independent of $n$. It should also be noted that in this regime, the key parameter $|t_{d^+}|^2+|t_{d^-}|^2-2|t_{d^0}|^2$ vanishes either at resonance, where the chain of atoms acts as a perfect mirror, or for $|\Delta /\gamma|=1$, where the four atomic configurations yield the same transmission coefficient for the photon, as illustrated in Fig.\,\ref{fig:partialTomo}. In the same figure, we find that the difference in the probability of transmission between the two extreme cases (no superposition and equal superposition) is largest when $|\Delta /\gamma|\approx 2.2$. Thus, the best resolution is achieved at this particular normalized detuning. While this version of the protocol requires less information on the position of the atoms, it is strongly dependent on the exact form of the trapping potential (\ref{eq:trappingPot}) (see Appendix \ref{sec:appB}). As a side remark, it should also be noticed that the periodicity of the trapping potential is essential in order to address the atoms identically, both during the initial state preparation and the Hadamard transformation.

\begin{figure}
\includegraphics[width=0.46\textwidth]{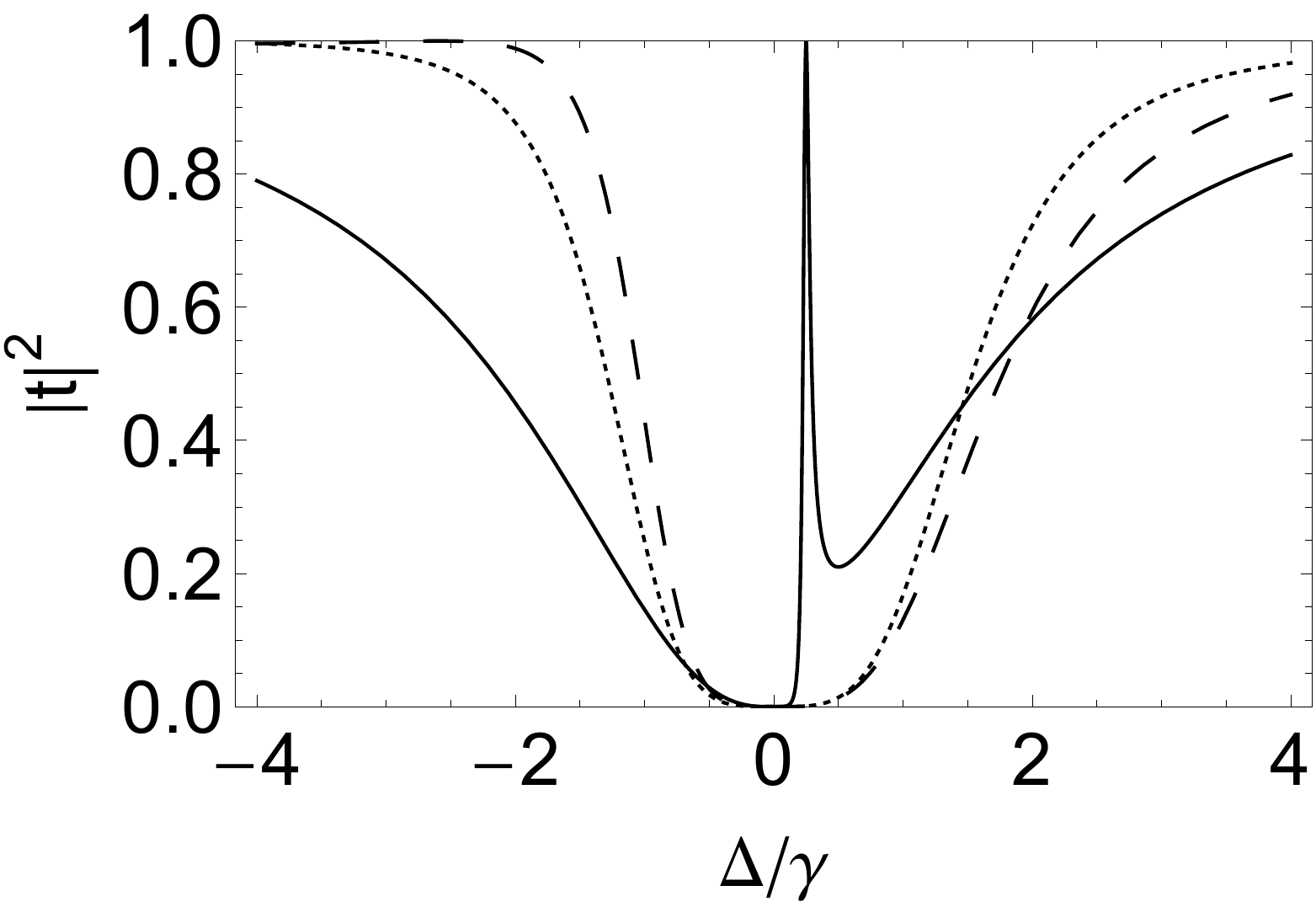}\label{fig:extractSpectra}
\caption{\label{fig:extractEnt}Transmission coefficient in a regime where extraction of a Bell state $|\Psi^+\rangle$ can be efficiently performed by probing at $\Delta /\gamma=0.25$, where $|t_{d^0}|=1$ and $|t_{d^\pm}|\approx 0$. The two atoms are in adjacent wells ($n=1$) and $2kl_t=(-\arctan(0.25)+\pi)$. The solid, dashed, and dotted line represent respectively the $d^0$, $d^+$, and $d^-$ configurations.}
\end{figure}

\begin{figure}
\includegraphics[width=0.46\textwidth]{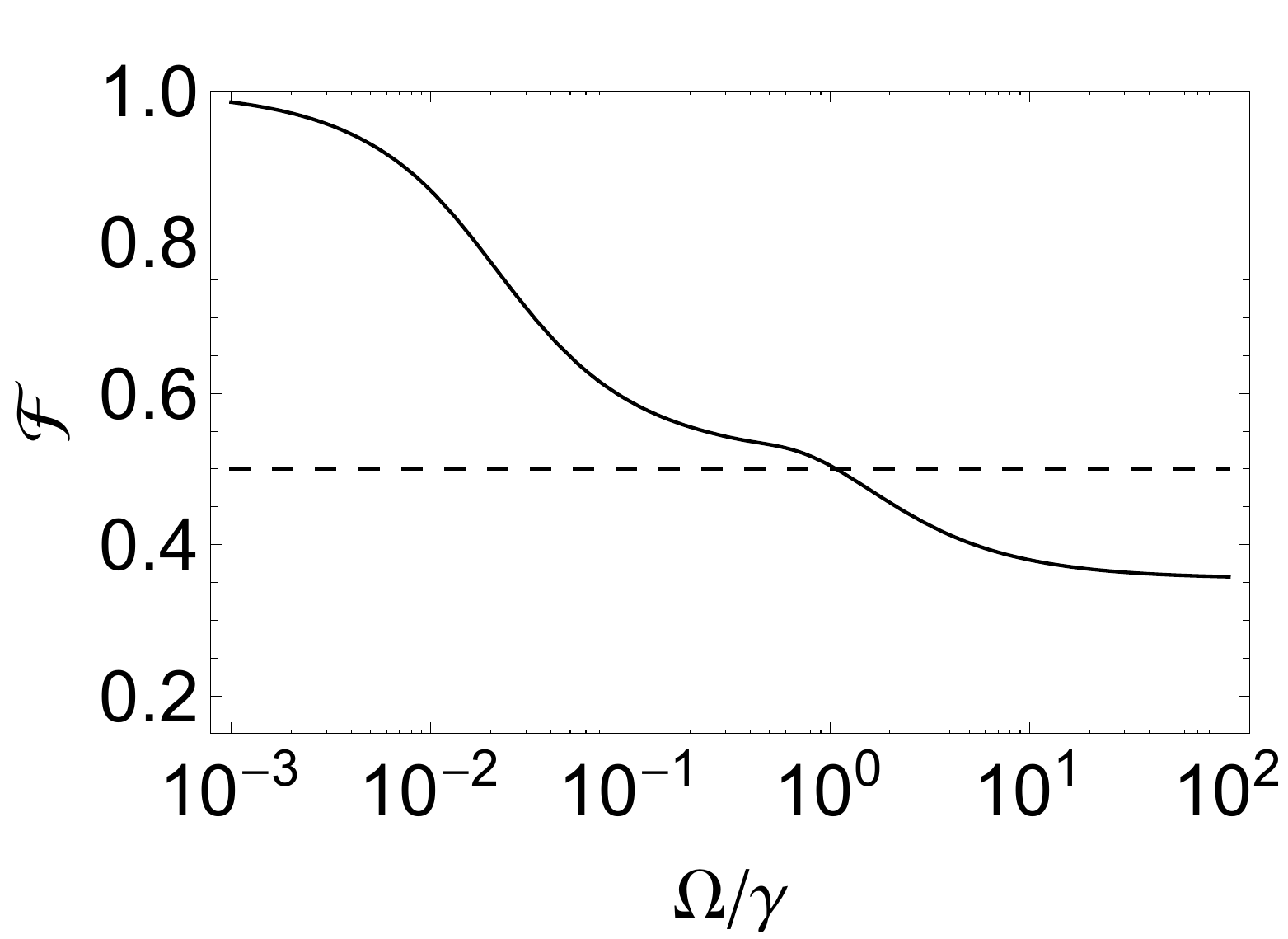}
\caption{\label{fig:fidelity} Singlet fidelity of the post-selected state (thick line) and the initial state $|\phi\rangle$ (dashed line) as a function of the ratio between the input-pulse bandwidth $\Omega$ and the interaction strength $\gamma$. Plotted for $|c_L|=1/\sqrt{2}$. Notice how if $\Omega\ll\gamma$, $\mathcal{F}=1$ can be achieved.}
\end{figure}

\section{Extracting entanglement}
In the remainder of this paper, we show how to herald a maximally entangled state from the atomic spatial state $|\phi\rangle$, by probing the transmitted photon. By contrast, in this last part the second atom is now more than a tool to obtain information on the superposition created. Given the form (\ref{eq:atomicSupState}) of $|\phi\rangle$, a judicious choice is to aim for the symmetric superposition $|\Psi^+\rangle=(|LR\rangle+|RL\rangle)/\sqrt 2$, such that the relative phase between the two spatial states is independent of $(c_L,c_R)$. We thereby look for trap parameters yielding a transmission peak for the states $|LL\rangle$ and $|RR\rangle$, which are defined for $\Delta \ne 0$ by the relation $2nkl_t=-\arctan(\Delta /\gamma)\pmod{\pi}$. In such a regime, the detection of a reflected photon projects the atomic spatial state onto $|\Psi^+\rangle$, up to a global phase. While a zero transmission for the configurations $d^\pm$ is optimal regarding the efficiency of the process, the only requirement is that it merely differs from 1. Figure \ref{fig:extractEnt} illustrates an almost optimal situation where a Bell state $|\Psi^+\rangle$ can be extracted by probing at $\Delta /\gamma=0.25$. The procedure depends on the actual superposition (\ref{eq:atomicSupState}) that is initially created only in its yield, that is $2|c_L c_R|^2$. Notice also that it requires more control on the loading of the atoms into the traps compared to the previous proposed protocol, since in this case $n$ has to be adjusted accordingly to the trap parameter $l_t$ in order to obtain a transmission peak at a fixed frequency.

The possibility of obtaining a maximally entangled state is restricted to the regime $\Omega\ll\gamma$. In other cases (see Appendix \ref{sec:appC}), the fidelity $\mathcal{F}(\rho)=\langle\Psi^+|\rho|\Psi^+\rangle$ of the post-selected state will be smaller than 1, as illustrated in Fig.\,\ref{fig:fidelity}. One should note that for an input pulse sufficiently sharp in the frequency domain, the fidelity is always enhanced by post-selecting on reflection events. Specifically, we find that the fidelity of the post-selected and initial state become equal when $\Omega=\gamma$.

\section{Conclusion}
We have shown how to obtain information on the coherence of an atomic spatial superposition, by sending a photon at a fixed frequency and studying the probability of transmission. Our scheme is thus well-suited to various applications in quantum information processing. We also considered entanglement extraction as an example of such applications. The theory analyzed here may form the basis for future experiments exploiting the one-dimensional waveguide set-up in the regime of strong light-matter interaction \cite{Hung2013,StrongCoup2,StrongCoup3}, in order to investigate and measure spatial superpositions. An exciting extension to the work presented in this paper would be to study how spatial superpositions of atomic mirrors affect cavity QED dynamics \cite{Chang2012}. We leave this for future investigations.

\begin{acknowledgements}
We thank Christian Kurtsiefer and Deborah S. Jin for fruitful discussions, and Feiyu Zhu for preliminary calculations. This research is supported by the National Research Foundation (partly through its Competitive Research Programme, Award No. NRF-CRPX-20YY-Z) and the Ministry of Education, Singapore.
\end{acknowledgements}

\appendix
\section{\label{sec:appA}Derivation of the transmission amplitude}
We give here a detailed derivation of the transmission coefficient (\ref{eq:transCoeffSF}) as given in the main text. For this, we first note that, within the single-excitation domain, any state of the system can be decomposed as:
\begin{eqnarray}
	|\psi (t)\rangle&=&\int_{0}^\infty\!\mathrm{d} \omega\, c_a(\omega,t)\hat{a}_\omega^\dagger |\varnothing\rangle +\int_{0}^\infty\!\mathrm{d} \omega\, c_b(\omega,t)\hat{b}_\omega^\dagger |\varnothing\rangle\nonumber\\ &+&c_{eg}(t) \hat{\sigma}^1_+|\varnothing\rangle+c_{ge}(t) \hat{\sigma}^2_+|\varnothing\rangle ,
\end{eqnarray}
where $|\varnothing\rangle\equiv|0\rangle_a|0\rangle_b|g\rangle_1|g\rangle_2$ corresponds to the forward and backward propagating modes being in vacuum state while both atoms are in the ground state. Specifically, the initial state consisting of a single-photon pulse incident from the left corresponds to $c_b(\omega,0)=c_{eg}(0)=c_{ge}(0)=0$ and $c_a(\omega,0)=f(\omega)$ where $f(\omega)$ is the shape of the photon pulse.

To describe the evolution of the system, it is convenient to work in a reference frame shifted with respect to the free Hamiltonian (first two terms in (\ref{eq:Hamiltonian})). The Schrödinger equation then reads (we omit the time dependence of the field and atoms variables for clarity)
\begin{eqnarray}\label{eq:SMSchroEq}
	&&\dot{\tilde{c}}_a(\omega)  = g_\omega\Big(c_{eg} +c_{ge} e^{-i\omega d/c}\Big)e^{i(\omega-\omega_A)t}\\
	&&\dot{\tilde{c}}_b(\omega)  = g_\omega\Big(c_{eg} +c_{ge} e^{i\omega d/c}\Big)e^{i(\omega-\omega_A)t}\nonumber\\
	&&\dot{c}_{eg}  = -\int_{0}^\infty\!\mathrm{d} \omega\, g_\omega \Big(\tilde{c}_a(\omega) +\tilde{c}_b(\omega) \Big)e^{-i(\omega-\omega_A)t} \nonumber\\
	&&\dot{c}_{ge}  = -\int_{0}^\infty\!\mathrm{d} \omega\, g_\omega \Big(\tilde{c}_a(\omega) e^{i\omega d/c}+\tilde{c}_b(\omega) e^{-i\omega d/c}\Big)e^{-i(\omega-\omega_A)t} . \nonumber
\end{eqnarray}
where $\tilde{c}_{a,b}(\omega) \equiv c_{a,b}(\omega)  e^{\pm i\omega x_1/c}$ is introduced to simplify the notation and $d=x_2-x_1>0$.

Formally integrating the field variables ${\tilde{c}}_a(\omega) $ and ${\tilde{c}}_b(\omega) $ and substituting them into the last two equations of (\ref{eq:SMSchroEq}), one gets a closed set of equations for the atomic variables under the Weisskopf-Wigner approximation \cite[p. 207]{QOptics1997}
\begin{eqnarray}\label{eq:SMclosedSet}
	\dot{c}_{eg}(t)  &=& -\gamma (c_{eg}(t)+c_{ge}(t-d/c)e^{i\omega_A d/c})\nonumber\\
	&&-\sqrt{\gamma} e^{-i\Delta t}\xi(t)\nonumber\\
	\dot{c}_{ge}(t)  &=& -\gamma (c_{ge}(t)+c_{eg}(t-d/c)e^{i\omega_A d/c})\nonumber\\
	&&-\sqrt{\gamma} e^{-i\Delta t}\xi(t-d/c)e^{i\theta_d} , 
\end{eqnarray}
where $\xi(t)\equiv \int_{0}^\infty\!\mathrm{d} \omega\, f(\omega) e^{-i(\omega-\omega_0)(t-x_1/c)}$. In the following we consider a square pulse
\begin{equation}
	\xi(t)=\left\{
    \begin{array}{cl}
        \sqrt{\frac{\Omega}{2}} \qquad& \text{for } 0\leq t\leq \frac{2}{\Omega} \\
        0\quad & \text{otherwise}
    \end{array}
\right. .
\end{equation}

\begin{figure*}
\subfloat{
\includegraphics[width=0.32\textwidth]{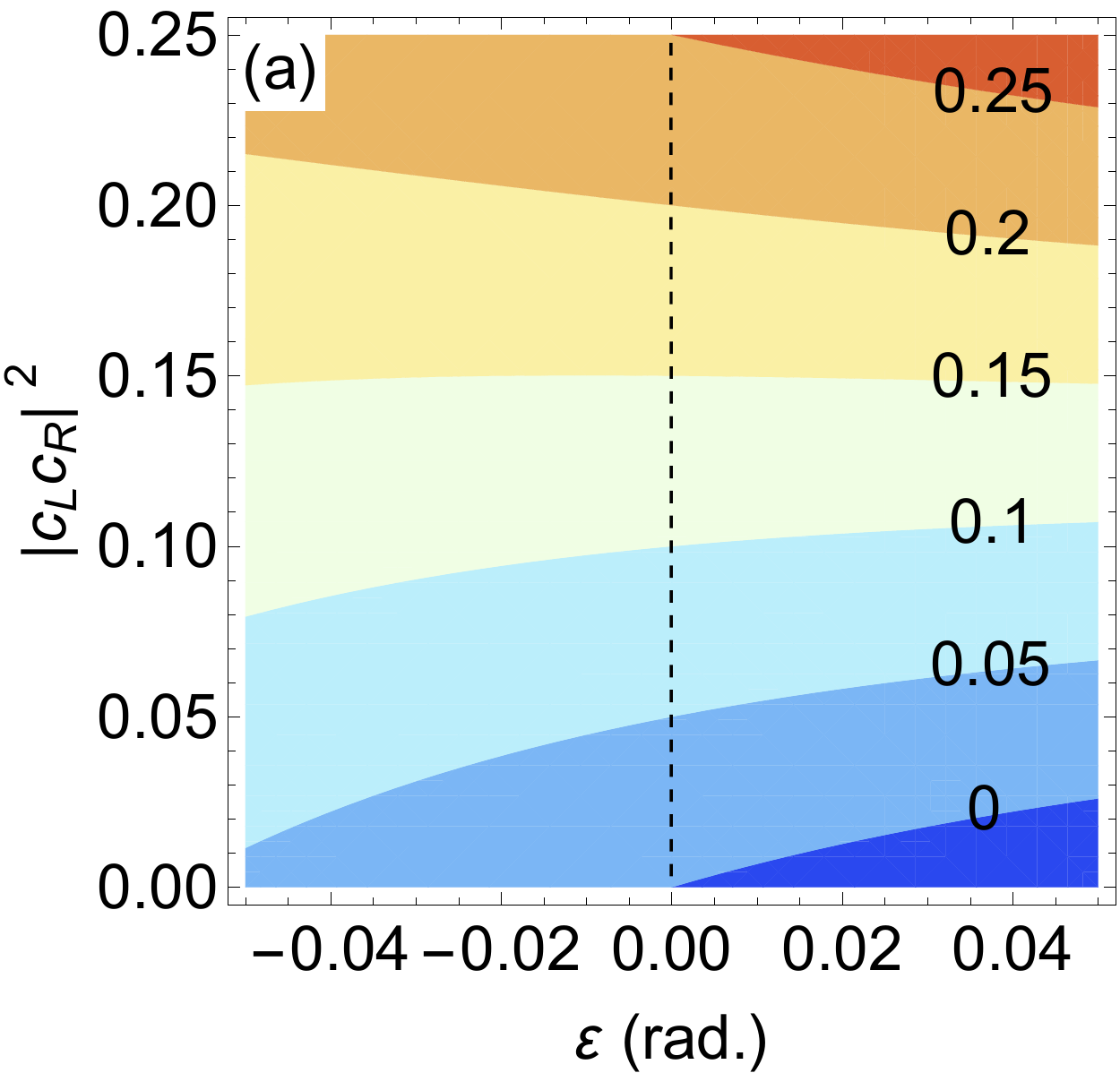}\label{fig:error1}
}
\subfloat{
\includegraphics[width=0.32\textwidth]{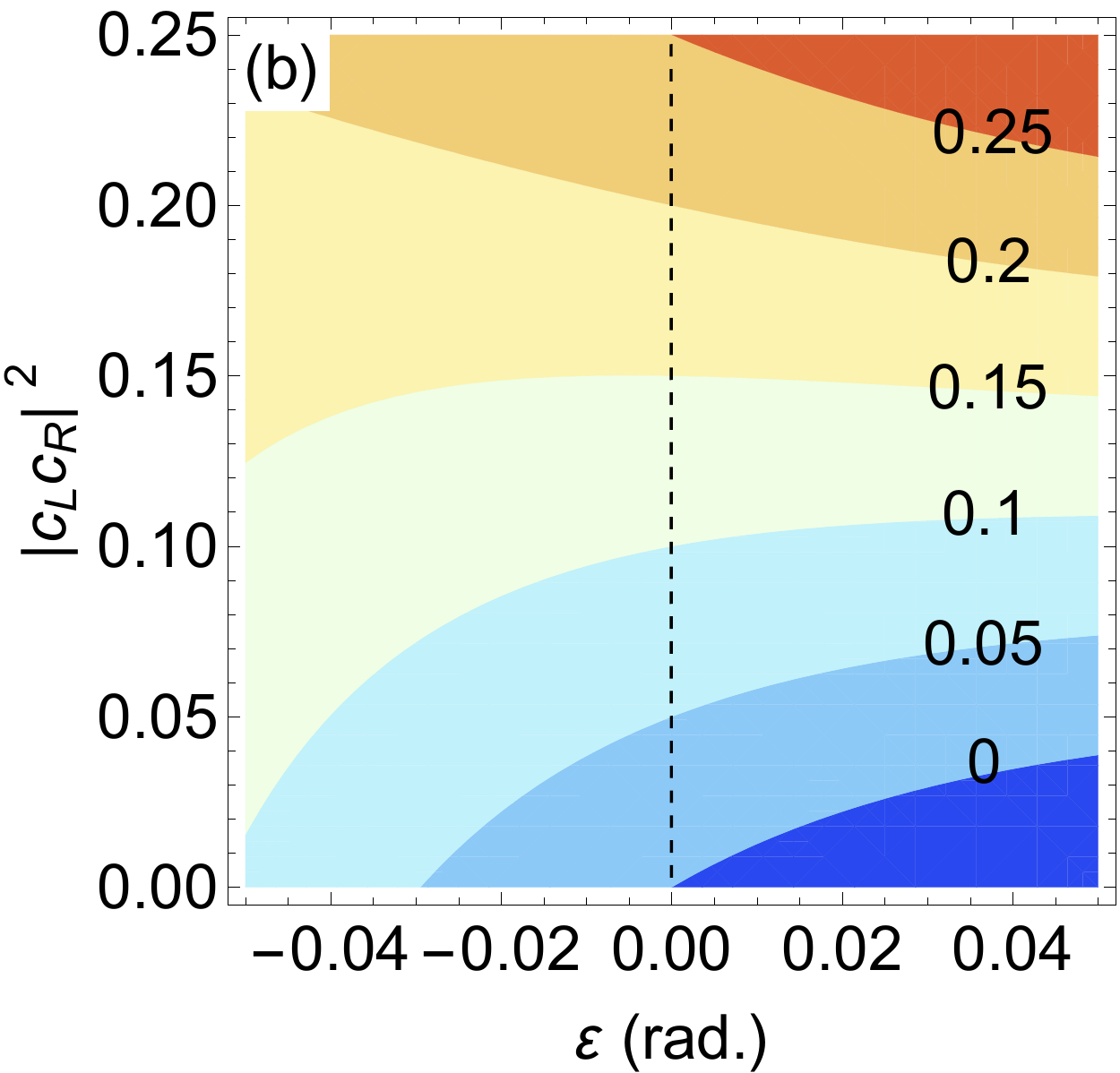}\label{fig:error2}
}
\subfloat{
\includegraphics[width=0.32\textwidth]{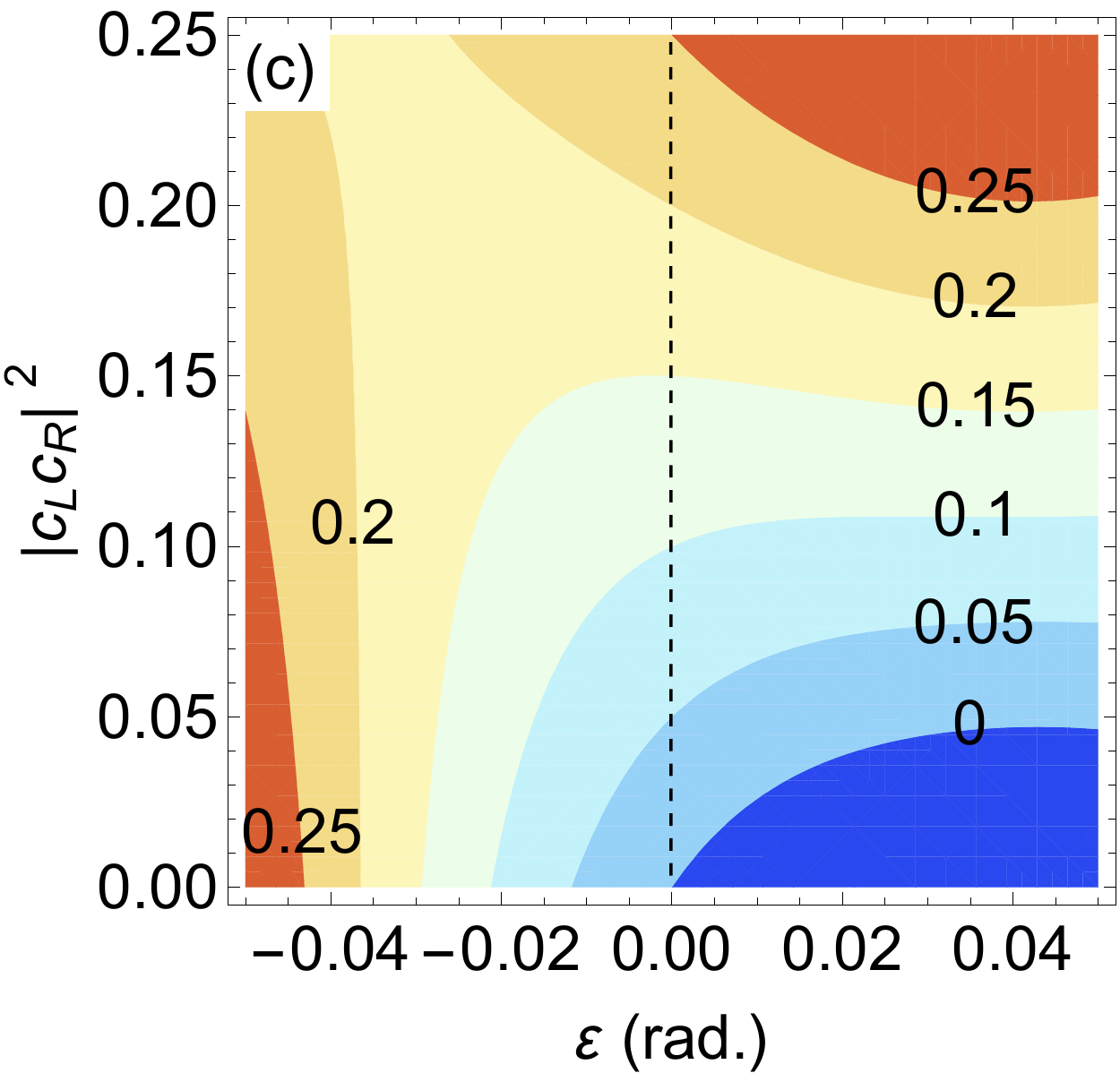}\label{fig:error5}
}
\caption{\label{fig:SM1}(color online).\quad Contour plot of $|c_Lc_R|^2$ derived from (\ref{eq:probTrans}) assuming the condition $kl_t=\pi /2$ is matched when computing the spectra $|t_{d^0}|$ and $|t_{d^\pm}|$. The probability of transmission $P_{|\phi\rangle}(\text{trans.})$, coming from an imperfect implementation, includes a correction term $\varepsilon\in\mathbb{R}$ such that $kl_t=\pi /2+\varepsilon$. The y-axis is the correct value of the superposition-weights product and the dashed line represents a perfect implementation for which the value derived from (\ref{eq:probTrans}) is correct. (a) The two atoms are in adjacent wells ($n=1$). (b) $n=2$. (c) $n=5$. For this last case, in the lower left corner, one would get $|c_Lc_R|^2\approx 1/4$ whereas there is no superposition in reality.}
\end{figure*}

Finally, solving the set of equations (\ref{eq:SMclosedSet}) allows us to obtain the transmission coefficient $|t_d|^2=1-\lim\limits_{t\to\infty} N_{\text{ref}}(t)$ with
\begin{eqnarray}\label{eq:SMNref}
	N_{\text{ref}}(t) &=& \int_{0}^\infty\!\mathrm{d} \omega\, \langle\psi (t)|\hat{b}_\omega^\dagger\hat{b}_\omega |\psi (t)\rangle\\
	&=& \int_{0}^\infty\!\mathrm{d} \omega\, |c_b(\omega)|^2 \nonumber\\
	&=& \gamma \int_{0}^{t}\!\mathrm{d} t^\prime\, |c_{eg}(t^\prime)+c_{ge}(t^\prime-d/c)e^{i\omega_A d/c}|^2 .\nonumber
\end{eqnarray}

When solving the set of equations (\ref{eq:SMclosedSet}) and then evaluating the integral (\ref{eq:SMNref}), one should note that we can neglect the time delay $d/c$ induced by the distance between the two atoms. This is justified because, under realistic experimental set-up,  $d/c$ is many order of magnitude smaller than the inverse of the interaction strength $1/\gamma$ which governs the timescale at which the system evolves. Then focusing on the near-resonant case ($\Delta\ll\omega_A$), we find that the transmission coefficient in the monochromatic limit ($\Omega \ll \gamma$) has the form of Eq.\,(\ref{eq:transCoeffSF}).

\section{\label{sec:appB}Sensitivity of the $n$-independent protocol}
In the main text we show that when the trap is designed such that the condition $kl_t=\pi /2 \pmod{\pi}$ is met, our protocol to obtain information on the atomic spatial superposition does not depend on $n$ anymore, which characterizes in which traps the atoms were initially loaded. However it then becomes very sensitive to the exact form of the trapping potential. This sensitivity is displayed in Fig.\,\ref{fig:SM1} where we introduce a correction term $\varepsilon\in\mathbb{R}$ modifying the trapping potential such that $kl_t=\pi /2+\varepsilon$. These results demonstrate that the error induced in the derivation of $|c_Lc_R|^2$ (\ref{eq:probTrans}) grows with the initial distance $n$ between the atoms.

\begin{figure}
\includegraphics[width=0.46\textwidth]{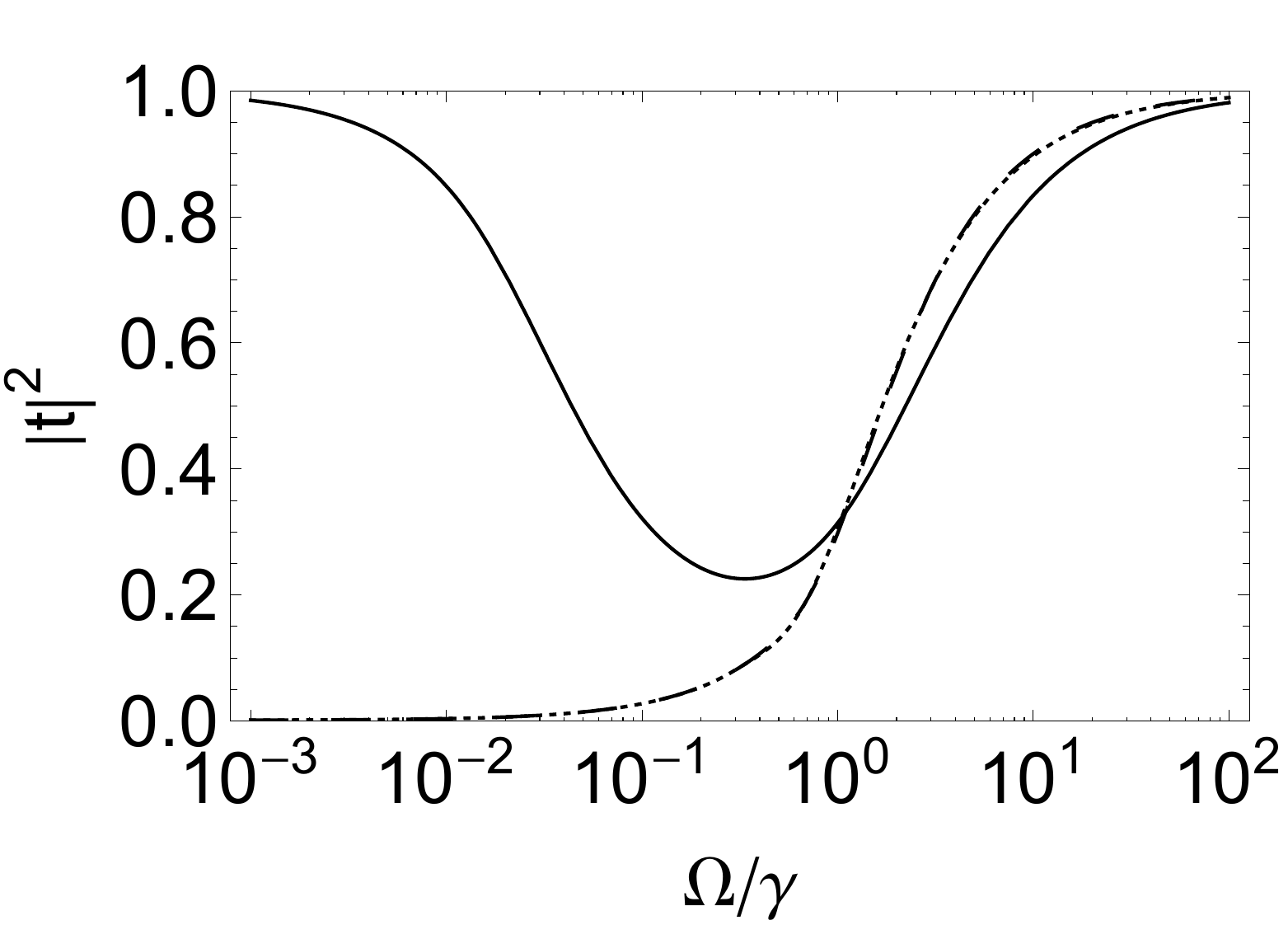}
\caption{\label{fig:SM1a} Transmission coefficient at $\Delta /\gamma=0.25$ as a function of the normalized bandwidth $\Omega / \gamma$ for the same parameters as in Fig.\,\ref{fig:extractEnt}. The solid, dashed, and dotted line represent respectively the $d^0$, $d^+$, and $d^-$ configurations.}
\end{figure}

\section{\label{sec:appC}Entanglement extraction in the non-monochromatic regime}

In the main text we propose a scheme to extract a singlet with fidelity $\mathcal{F}_{|\Psi^+\rangle}=1$ from the state $|\phi\rangle$ (\ref{eq:atomicSupState}). This is achieved by post-selecting on reflection events at a frequency where the configuration $d^0$ leads to a transmission peak. Here we extend these results beyond the monochromatic limit. Specifically, Fig.\,\ref{fig:SM1a} illustrates how the transmission spectra varies at the peak frequency as a function of the normalized input-pulse bandwidth $\Omega / \gamma$. For $\Omega \ll \gamma$, we find back the results obtained in the monochromatic limit (see Fig.\,\ref{fig:extractEnt}). On the other hand, we obtain full transmission for all the configurations when $\gamma \ll \Omega$. This result highlights the importance of the ratio $\Omega / \gamma$ when exploiting light-matter interaction in one-dimensional waveguides. Finally, we observe that when $\gamma<\Omega$, the probability of transmission in the configuration $d^0$ is the smallest. This is in agreement with Fig.\,\ref{fig:fidelity} where the fidelity of the post-selected state is compared to the initial state's fidelity $\mathcal{F}(|\phi\rangle\langle\phi |)=\sqrt{2}|c_L c_R|$ as a function of $\Omega / \gamma$. Indeed, as illustrated, the fidelity is increased by post-selecting only when $\Omega<\gamma$. This is due to the fact that when $\gamma<\Omega$, the probability of reflection is higher in the configuration $d^0$ than in $d^\pm$.

\bibliography{Bibliography}

\end{document}